\documentclass[%
 reprint,
superscriptaddress,
 amsmath,amssymb,
 aps,
prb,
]{revtex4-2}

\usepackage{tikz}
\usetikzlibrary{arrows.meta,positioning,decorations.pathmorphing,decorations.markings}
\tikzset{>=Stealth}
\tikzset{->-/.style={decoration={markings,mark=at position {0.5} with {\arrow{Stealth}}},postaction={decorate}}}
\usetikzlibrary{decorations.pathreplacing}

\usepackage{gensymb}
\usepackage{graphicx}% Include figure files
\usepackage{dcolumn}% Align table columns on decimal point
\usepackage{bm}% bold math
\usepackage{hyperref}% add hypertext capabilities
\hypersetup{colorlinks=true, citecolor = blue, linkcolor = blue, urlcolor = blue}
\usetikzlibrary{shapes.multipart}
\usepackage{lipsum} 

\begin{document}

\preprint{APS/123-QED}

\title{Realization of giant elastocaloric cooling at cryogenic temperatures in TmVO$_4$ via a strain load/unload technique }% Force line breaks with \\

\author{Mark P. Zic}
\thanks{zic@stanford.edu}
\affiliation{Geballe Laboratory for Advanced Materials, Stanford University, Stanford, CA 94305}
\affiliation{Department of Physics, Stanford University, Stanford, CA 94305}
\author{Linda Ye}
\affiliation{Geballe Laboratory for Advanced Materials, Stanford University, Stanford, CA 94305}
\affiliation{Department of Applied Physics, Stanford University, Stanford, CA 94305}
\author{Maya H. Martinez}
\affiliation{Geballe Laboratory for Advanced Materials, Stanford University, Stanford, CA 94305}
\affiliation{Department of Applied Physics, Stanford University, Stanford, CA 94305}
\author{Ian R. Fisher}
\affiliation{Geballe Laboratory for Advanced Materials, Stanford University, Stanford, CA 94305}
\affiliation{Department of Applied Physics, Stanford University, Stanford, CA 94305}

\date{\today}% It is always \today, today,

\begin{abstract}
%\lipsum[1-2]

The adiabatic elastocaloric effect relates changes in the strain that a material experiences to resulting changes in its temperature. While elastocaloric materials have been utilized for cooling in room temperature applications, the use of such materials for cryogenic cooling remains relatively unexplored. Here, we use a strain load/unload technique at low temperatures, similar to those employed at room-temperature, to demonstrate a large cooling effect in TmVO$_4$. For strain changes of $1.8 \cdot 10^{-3}$, the inferred cooling reaches approximately 50\% of the material's starting temperature at 5 K, justifying the moniker ``giant". Beyond establishing the suitability of this class of material for cryogenic elastocaloric cooling, these measurements also provide additional insight to the entropy landscape in the material as a function of strain and temperature, including the behavior proximate to the quadrupolar phase transition.

\end{abstract}

\maketitle

\section{Introduction}

Low temperature cooling in the Kelvin and sub-Kelvin regimes is of considerable interest because of its application in low temperature research and, more recently, for efficient cooling of quantum computers. The workhorse method of dilution refrigeration relies on the availability of $^3$He, a scarce and costly resource, as well as requiring complex gas handling systems. Magnetocaloric cooling is simpler, but requires large magnetic fields, which can be incompatible with certain applications and cannot be cycled rapidly. Alternate methods of cryogenic cooling are consequently of particular interest. Until recently, elastocaloric cooling in the cryogenic regime was largely unexplored, in part because methods to measure the elastocaloric effect at cryogenic temperatures had not yet been developed, and in part because of the lack of candidate materials. Despite these challenges, elastocaloric cryogenic cooling using piezoelectric actuators offers several distinct advantages, including fast response times, localized cooling, and small space requirements. The elastocaloric effect can also provide unique insights into the entropy landscape and phases present in quantum materials, further motivating the development of new low-temperature techniques and methodologies, in addition to exploring candidate materials with suitably large responses for cryogenic applications  \cite{Straquadine2022-PRX,Ikeda2021-ECEB2gFluc,Li2022-MackenzieECE,Ye2023-DyB2C2,Gati2023-YbPtBi,Rosenberg2024-TmAg2}.

One of the prerequisites for elastocaloric cooling is a material for which isothermal changes in strain yield a large change in entropy. Consequently, materials that exhibit a low-temperature cooperative Jahn-Teller effect are suitable candidates \cite{Zic2024-TmVO4} (although other mechanisms are also possible \cite{Strassle2002-Ce3Pd20Ge6ECE}). Recently, TmVO$_4$, a well-known low temperature Jahn-Teller material, has been experimentally established to possess a large change in entropy with respect to strain, and hence a giant elastocaloric effect \cite{Zic2024-TmVO4}. For this reason, in this manuscript, TmVO$_4$ is used as an archetypal material to realize, for the first time, a large elastocaloric cooling at low temperatures via such a mechanism. 

TmVO$_4$ is an insulator. It undergoes a second-order, $xy$ symmetric ferroquadrupolar phase transition at $T_Q = 2.15$ K \cite{Melcher1976-Review,Gehring1975-review}. This electronic order accompanies a tetragonal ($T>T_Q$, space group $I4_1/amd$) to orthorhombic ($T<T_Q$, space group $Fddd$) \cite{Jiang2022-thesis} structural phase transition; this coupled electronic - structural instability is the cooperative Jahn-Teller effect, and TmVO$_4$ has long been seen as an archetype \cite{Melcher1976-Review,Gehring1975-review}. In the tetragonal phase, the crystal electric field (CEF) groundstate of the Tm$^{3+}$ ion is a non-Kramers doublet \cite{Knoll1971-CEFspectra}. The associated quadrupole moment couples linearly to strain with the same symmetry, such that the doublet splits spontaneously in the ordered state, releasing $k_B \ln{2}$ of entropy per Tm ion \cite{Cooke1972-ortho,Li2024-TmYVO4,Massat2022-PNAS}. Similarly, for temperatures above $T_Q$, the doublet splits linearly in response to applied strain. This enables direct control of a large reservoir of entropy above $T_Q$, $k_B \ln{2}$ per Tm ion, using $xy$ symmetric strain. Therefore, under adiabatic conditions, the temperature of TmVO$_4$ can be precisely controlled by externally driven, anti-symmetric strain. Also, the relevant elastic mode, $c_{66}$, is incredibly soft at low temperatures (16 GPa \cite{Melcher1973-elastic-constants}) and softens significantly as the phase transition is approached, reducing risk of sample failure. Overall, TmVO$_4$ is an excellent material candidate for initial elastocaloric cooling efforts.

Techniques to measure the low temperature elastocaloric response of materials have only recently been developed. Thus far, an AC technique has been employed, in which the strain is oscillated over a small range, at a frequency which is high enough to ensure thermal decoupling between the sample and the apparatus used to induce the strain, thus permitting measurement of the adiabatic response \cite{Ikeda2019-ECEtechnique}. Typical AC strains are of order 10$^{-5}$, and typical temperature changes are of order 1 mK. By varying the DC bias strain, the entropy as a function of strain and temperature can be extracted using this technique; this is how we know that TmVO$_4$ is a suitable candidate \cite{Zic2024-TmVO4}. However, to demonstrate significant elastocaloric cooling at low temperatures requires a method that more closely resembles that which would be used for cooling purposes, with much larger changes in strain and temperature. Akin to processes employed at room temperature \cite{Chauhan2015-eceReview,Engelbrecht2019-eceDeviceReview}, the method utilized in the present manuscript is a strain load/unload (or pulse) technique. Beyond demonstrating substantial cooling, this load/unload technique also allows for measurements and analyses in the time domain, in contrast to the AC elastocaloric effect, which measures in the frequency domain; this allows for the experimental capture of new effects, especially those associated with the phase boundary of the candidate material. In terms of hardware, the setup closely resembles those that have been implemented for the AC technique that we first developed \cite{Ikeda2019-ECEtechnique} (see Appendix A). Because of the similarity between AC strain measurement setups and the one presented here, results from the two techniques can be compared for the same sample with the same mounting, providing key insights into the associated thermal time constants and how they manifest in each measurement (see Appendix B).

\section{Experimental Methods}

Single crystals of TmVO$_4$ were grown via a flux technique \cite{Feigelson1968-Growth,Smith1974-Growth}. Samples were cut and polished to render long, thin rectangular parallelepiped bars, with the long axis along the tetragonal [110] crystallographic direction. Uniaxial stress applied along that direction yields a combination of $xy$-symmetry shear strain, $\varepsilon_{xy}$, together with a small amount of symmetry preserving strains, $(\varepsilon_{xx}+\varepsilon_{yy})/2$ and $\varepsilon_{zz}$ \cite{Ikeda2018-strainChannels}. Because $c_{66}$ is soft compared to $\frac{1}{2}(c_{11}+c_{22})$, nearly all of the strain (approximately 90\%, which varies slightly as a function of temperature) is $xy$ strain (see Appendix A and supplementary material of ref. \cite{Zic2024-TmVO4} for more details). We adopt a coordinate system in which unprimed coordinates refer to the [100], [010], and [001] crystallographic directions, and primed coordinates refer to the stress direction, rotated by 45 degrees about the z-axis (see Appendix A). Strain in the $x'$ direction (i.e. along [110]), $\varepsilon_{x' x'}$, was measured in the experiment, and is reported in figures including strain transfer corrections (see Appendix A). The material was strained using a Razorbill CS100 strain cell. The sample was suspended between two pairs of sample plates to which it was epoxied. Piezoelectric stacks controlled the strain that the material experienced, and the temperature of the sample was measured using a RuO$_2$ thin film thermometer. The strain that the sample experiences was measured using a capacitive sensor integrated on the strain cell. More details on the experimental setup can be found in Appendix A.

\section{Results}

\begin{figure}[thbp]
\includegraphics[width=\linewidth]{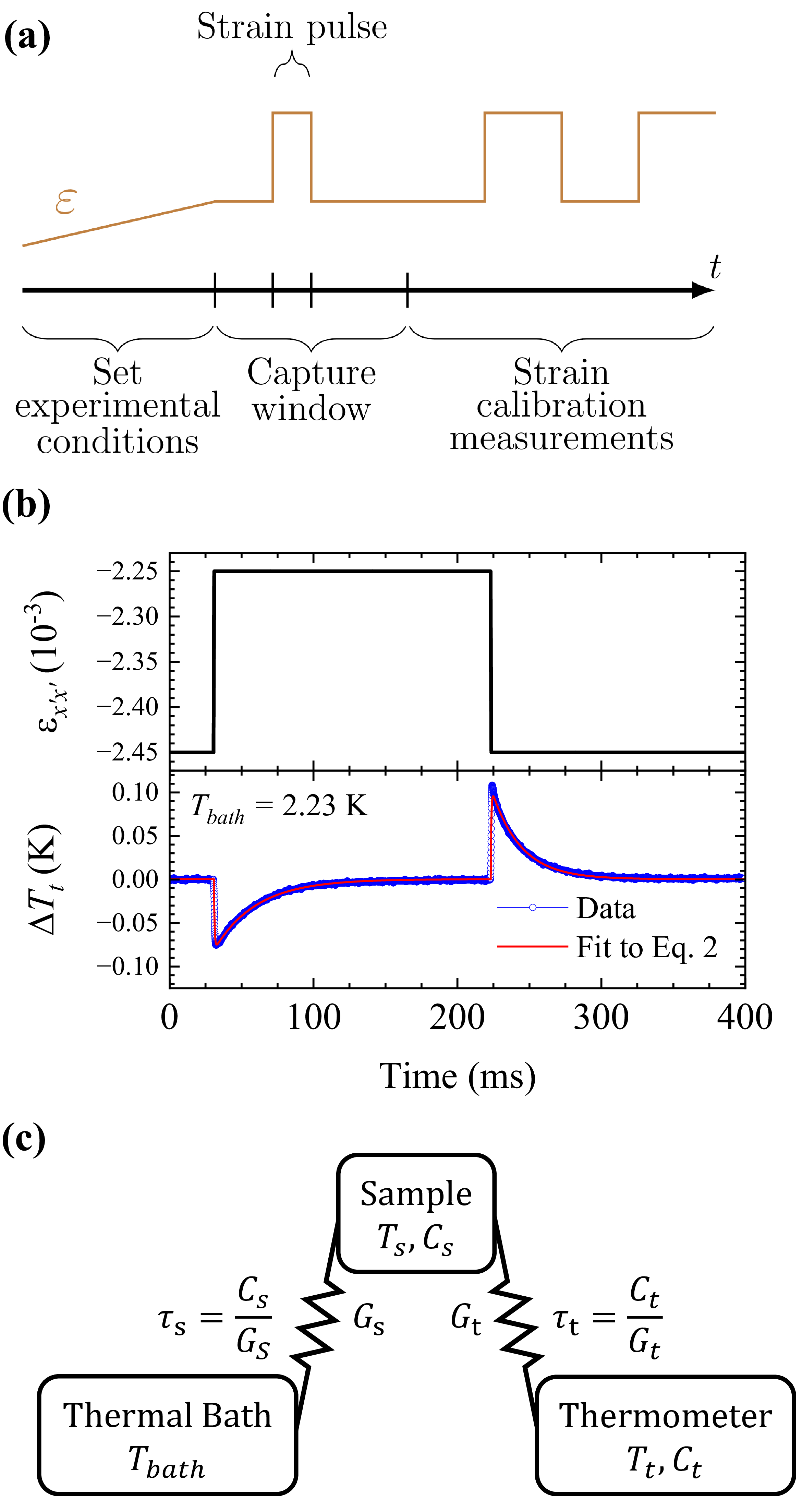}
\caption{\emph{}(a) A workflow timeline of the measurements, with the strain represented by the brown line, not to scale. The $x$-axis represents the flow of time during the measurement and is labeled with the different stages of the experiment, not to scale. (b) Representative data (upper panel) of the approximate strain experienced by the sample plotted alongside (lower panel) the temperature of the sample relative to the bath read out via an attached thermometer, both as a function of time. In the lower panel, blue points represent experimental data and the red line represents a fit ($R^2 > 0.99$) to the model described by Eq. 2.(c) The temperature of the thermometer ($T_t$) and sample ($T_s$) can be accurately described by a simple heat flow model governed by two time constants ($\tau_t$ and $\tau_s$, thermometer and sample time constant respectively, in the figure). $C_s$ and $C_t$ are the thermometer and sample heat capacities, respectively; $G_s$ and $G_t$ are the thermometer and sample thermal conductances, respectively. The thermal connections between thermal bath and thermometer to the sample are represented by thermal resistors.}
\end{figure}

Fig. 1(a) shows the measurement workflow. Before each measurement, the temperature and strain of the sample are set. Then, the capture begins and the strain pulse is subsequently sent. After the capture concludes, the pulse width is adjusted to a much longer time (on the order of one second) and the change in strain is measured as a function of time over several cycles to make a reliable determination. The process is then repeated for each measurement. Fig. 1(b) shows a representative pulse and response, with both cooling and heating curves, where the rise time of the strain (500 $\mu$s) is much shorter than the time constants associated with thermal relaxation. The focus of the remainder of this manuscript will be the signature of cooling.

Assuming that the time variation in the temperature of the material is solely due to thermal conduction to the bath, a simple thermal model can be used \cite{Ikeda2019-ECEtechnique, Strassle2002-Ce3Pd20Ge6ECE}. Fig. 1(c) displays a schematic of this model. The model gives the temperature of the sample as:
\begin{equation}
    T_{s}(t) = T_{bath} + \Delta T_s e^{-\frac{t}{\tau_{s}}},
\end{equation}
and gives the temperature of the thermometer as:
\begin{equation}
    T_{t}(t) = T_{bath} + \frac{\tau_{s}}{\tau_{t} - \tau_{s}}\Delta T_s \Big(e^{-\frac{t}{\tau_{s}}} + e^{-\frac{t}{\tau_{t}}}\Big).
\end{equation}
Here, $T_{s}$ is the temperature of the sample, $T_{t}$ is the temperature of the thermometer, $\tau_{s}$ ($C_s/G_s$) is the time constant between the sample and the bath, $\tau_{t}$ ($C_t/G_t$) is the time constant between the thermometer and the sample, $\Delta T_s$ is the intrinsic change in temperature of the sample caused by the change in strain, $C_s$ and $G_s$ are the heat capacity and thermal conductance of the sample, and $C_t$ and $G_t$ are the heat capacity and thermal conductance of the thermometer, respectively. Necessarily, $C_s >> C_t$. The representative curve shown in Fig. 1(b) fits well to Eq. (2) with an excellent $R^2$ value of 0.997.

The elastocaloric response can be obtained across the $\varepsilon_{xy}-T$ landscape to track systematic changes as a function of each independent variable. For example, the elastocaloric cooling response at 4 K with a strain pulse of $2 \cdot 10^{-4}$ for different values of the initial strain $\varepsilon_{x' x'}^{(0)}$ is shown in Fig. 2(a). As the magnitude of compressive strain that the sample experiences increases, the cooling of the sample initially increases, then decreases once the strain goes beyond $\varepsilon_{x' x'}^{(0)} = -1.8 \cdot 10^{-3}$. The time constant associated with thermal relaxation from the sample to the bath increases slightly as compressive strain is increased.

\begin{figure}[!t]
\includegraphics[width=\linewidth]{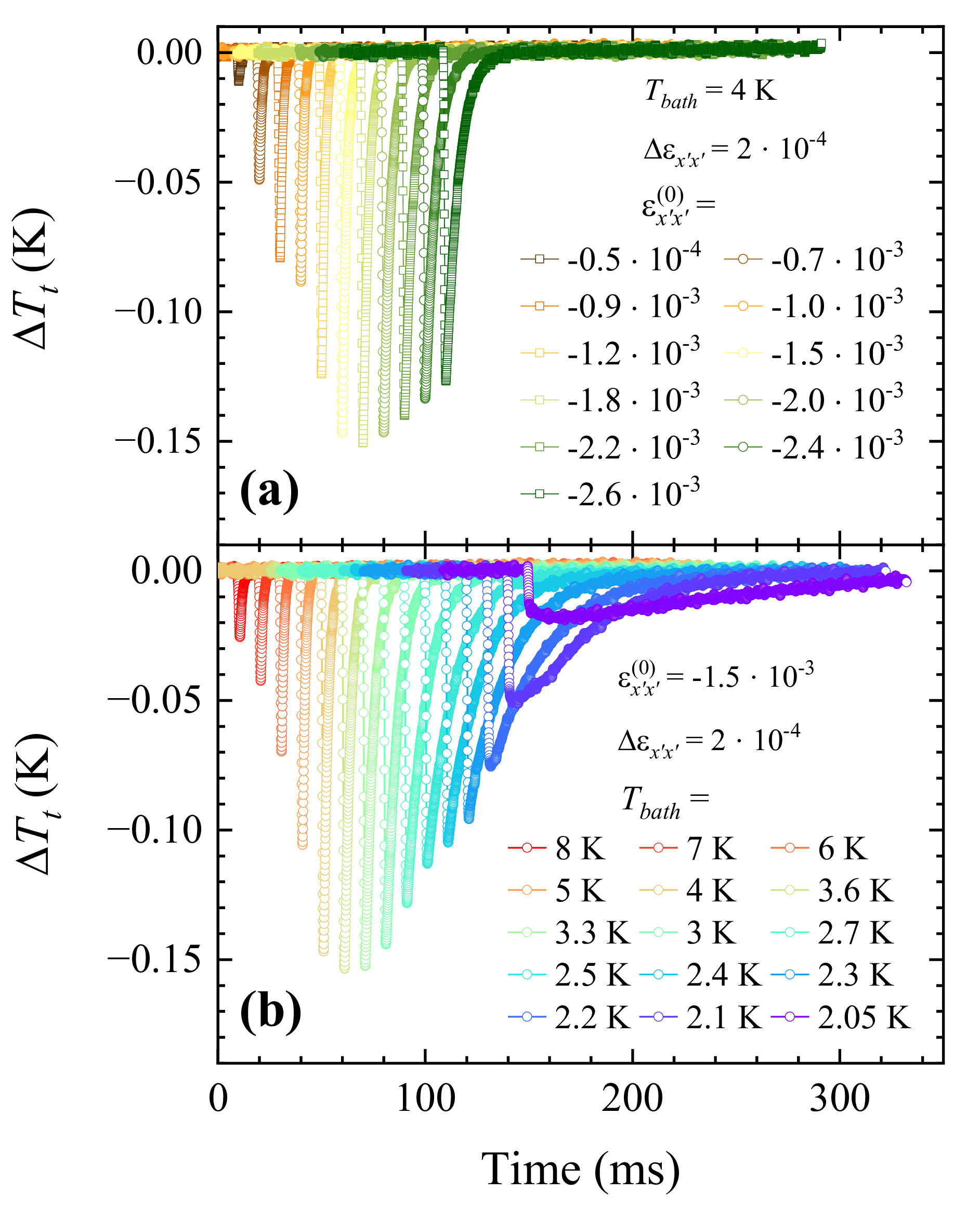}
\caption{ Curves are offset horizontally for clarity. (a) Representative elastocaloric response as measured by the thermometer ($\Delta T_t$) as a function of time (in milliseconds) under a moderate strain pulse ($\Delta \varepsilon_{x'x'} = 2 \cdot 10^{-4}$) for different initial strains ($\varepsilon_{x'x'}^{(0)}$) at $T_{bath} = 4$ K. (b) Representative elastocaloric response as measured by the thermometer ($\Delta T_t$) as a function of time (in milliseconds) under a moderate strain pulse ($\Delta \varepsilon_{x'x'} = 2 \cdot 10^{-4}$) for different bath temperatures ($T_{bath}$) with a starting strain of $\varepsilon_{x'x'}^{(0)} = -1.5 \cdot 10^{-3}$.}
\end{figure}

A more drastic change in the elastocaloric response is evident when temperature is varied, as shown in Fig. 2(b). While at high temperatures relative to the phase transition temperature (8 K), the elastocaloric response is small and grows considerably until it reaches a maximum at approximately 3.6 K, which agrees with previous AC elastocaloric measurements \cite{Zic2024-TmVO4}. As the sample is cooled further, the elastocaloric response begins to shrink and the thermal relaxation time of the sample increases dramatically, with significant temperature changes lasting for hundreds of milliseconds.

In an effort to achieve the largest temperature changes possible, the strain pulse amplitude was increased nearly ten fold to approximately $1.8 \cdot 10^{-3}$. The elastocaloric response was then measured at different temperatures, and the results are displayed in Fig. 3(a). Once again, the response is initially small and increases as the bath temperature decreases, albeit here the response peaks at approximately 5 K. The peak realized temperature change of the thermometer is just over 900 mK. As the bath temperature is further cooled, the elastocaloric response decreases, though a clear shoulder-like feature appears, deviating from the functional form predicted by the heat flow model given in Eq. 2. The initial, sharp increase in temperature for each curve is labeled with a black star, and the end of the shoulder-like feature for each curve is labeled by a red star. The feature is shown in all curves from 4 K down to 2.4 K and grows in size considerably. This feature is even subtly present in Fig. 2(b) in the data obtained at 2.1 K.

The failing of Eq. 2 originates from the absence of a time or temperature dependence of the thermal time constants in the model. This is most pertinent for the sample time constant ($\tau_s$), as the specific heat\cite{Massat2022-PNAS,Cooke1972-ortho,Li2024-TmYVO4} and thermal conductivity \cite{Daudin1982-thermalConductivity,Vallipuram2024-thermalCond} of TmVO$_4$ vary dramatically as a function of temperature near the phase transition, with the former varying more of the two. The specific heat of TmVO$_4$ grows significantly as the temperature is lowered to $T\approx T_Q$, although the exact temperature dependence depends on the strain state of the material \cite{Massat2022-PNAS,Cooke1972-ortho,Li2024-TmYVO4}, while the thermal conductivity has a small but noticeable kink \cite{Daudin1982-thermalConductivity,Vallipuram2024-thermalCond}. 

As a visual aid, the process of elastocaloric cooling in TmVO$_4$ using the strain load/unload technique is shown in Fig. 3(b) using the calculated entropy landscape of TmVO$_4$ \cite{Zic2024-TmVO4}. Here, for simplicity, we only show the entropy as a function of one component of the strain tensor, $\varepsilon_{xy}$; this component dominates the total strain value (see Appendix A), but nevertheless this figure should only be used for qualitative comparison with experimental data, for the total temperature change when symmetric strains are also present. Under an adiabatic change in strain, the material follows an isentrope, which will dictate its change in temperature. As the material cools via its own elastocaloric effect, it enters the ordered state, increasing its thermal time constant dramatically. In addition, in the ordered state, the entropy of the material does not change as a function of applied strain, except due to changes arising from domain walls. Hence, the cooling effect stops, and the isentrope in Fig. 3(b) becomes vertical as the material enters the ordered state. Once the material is in this ordered state and the change in strain has ceased, heat leaks into the material from the environment. The deeper the material is in the ordered state (i.e., the left white arrow is deeper in the ordered state compared to the right white arrow in Fig. 3(b)), the longer it will be in the ordered state, and the longer it will have a larger time constant, reinforcing the effect. In Fig. 3(b), this occurs because one elastocaloric process starts at a lower temperature than the other. This agrees with Fig. 3(a) in which the shoulder-like feature persists for a longer time when the sample is at a lower temperature, indicating that the sample takes a longer time to warm into the unordered state. A more quantitative representation of this is present in Fig. 3(c), the estimated sample time constant from specific heat (see supplementary material of ref. \cite{Zic2024-TmVO4}) and thermal conductivity data \cite{Daudin1982-thermalConductivity}. This quantity increases by four orders of magnitude from 8 K to 2 K, the sharpest increase being near the phase transition, bolstering the notion that the shoulder in Fig. 4(a) originates from a dramatic change in the time constant of the sample.

\begin{figure}[thbp]
\includegraphics[width=\linewidth]{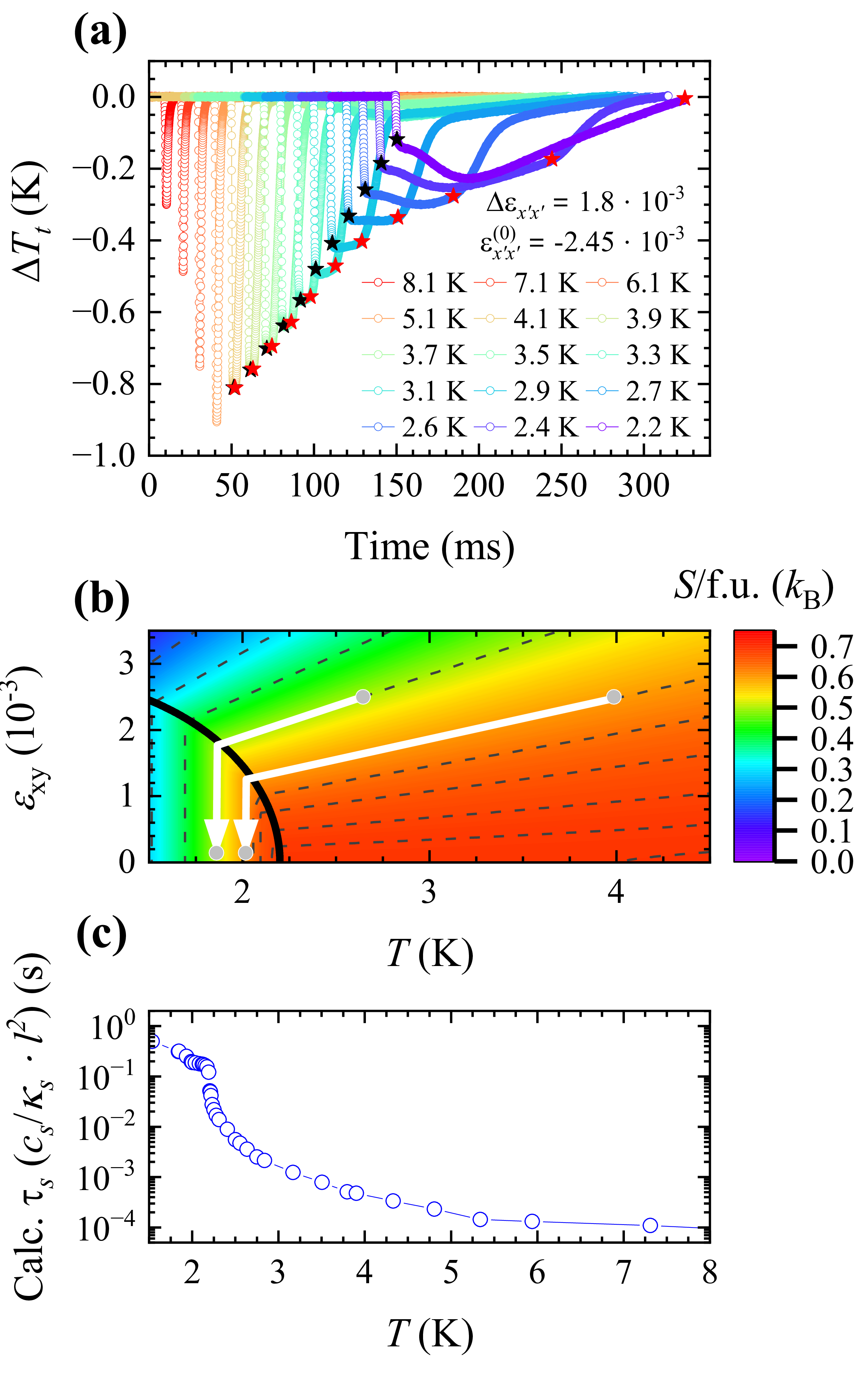}
\caption{(a) Representative elastocaloric response as measured by the thermometer ($\Delta T_t$) as a function of time (in milliseconds) under a large strain pulse ($\Delta \varepsilon_{x'x'} = 1.8 \cdot 10^{-3}$) for different bath temperatures ($T_{bath}$) with a initial strain of $\varepsilon_{x'x'}^{(0)} = -2.45 \cdot 10^{-3}$ (curves are offset horizontally for clarity). Black stars label the entrance to, and red stars label the exit from, the plateau region. The small, broad feature is due to imperfect background subtraction (see Appendix C). (b) The calculated entropy (color scale) of TmVO$_4$ \cite{Zic2024-TmVO4} as a function of $\varepsilon_{xy}$ and $T$. Dashed lines represented paths of constant entropy. Starting from the same strain and undergoing the same change in strain, the white arrows (isentropic paths) demonstrate that the sample will be deeper in the ordered phase if the initial (bath) temperature is lower. As heat leaks back into the sample from the sample environment, the sample initially remains in the ordered state, rearranging domains because the sample is held at finite strain (rather than stress). Once the sample is warmed to the unordered phase, thermal relaxation with a smaller $\tau_s$ begins. The entropy landscape is even with respect to $\varepsilon_{Xy}.$(c) The temperature dependence of $c_s/\kappa_s \cdot l^2$, where $c_s$ (see supplementary of material of ref. \cite{Zic2024-TmVO4}) and $\kappa_s$ \cite{Daudin1982-thermalConductivity} are both found in the literature and $l$ is the sample length, represents an estimation of the thermal time constant of the sample. }
\end{figure}

\begin{figure}[!t]
\includegraphics[width=\linewidth]{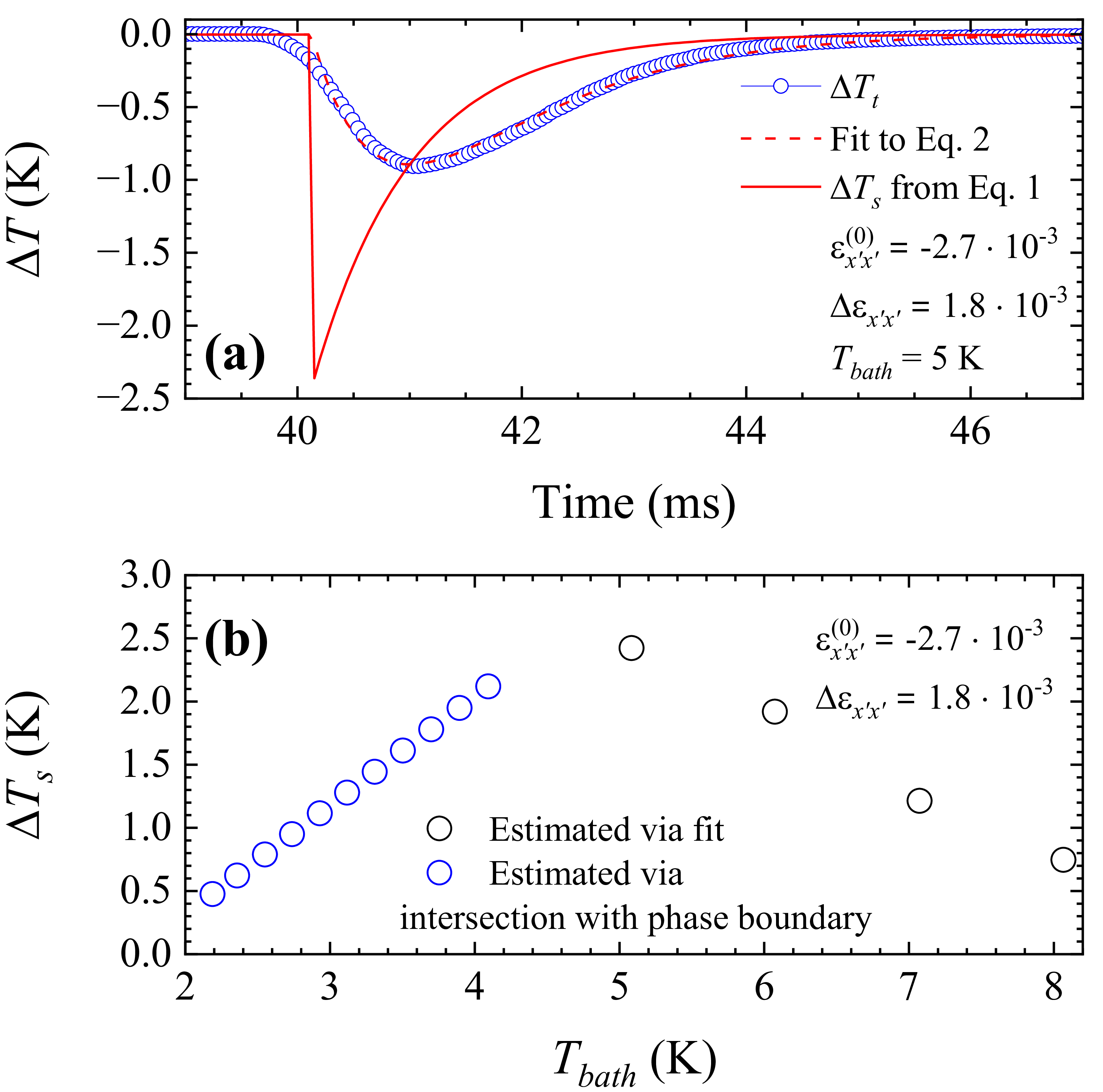}
\caption{(a) The temperature of the thermometer as a function of time for a representative, giant elastocaloric response (blue points), with an initial strain of $\varepsilon_{x'x'}^{(0)} = -2.7 \cdot 10^{-3}$ and a change in strain of $\Delta \varepsilon_{x'x'} = 1.8 \cdot 10^{-3}$ at $T_{bath} = 5\ \mathrm{K}$, plotted alongside the fit to Eq. 2 (dashed red) and the estimated sample temperature found using Eq. 1 (solid red). The estimated cooling is approximately 2.36 K, nearly half of the initial sample temperature. (b) The change in sample temperature ($\Delta T_s$), with an initial strain of $\varepsilon_{x'x'}^{(0)} = -2.7 \cdot 10^{-3}$ and a change in strain of $\Delta \varepsilon_{x'x'} = 1.8 \cdot 10^{-3}$, as a function of bath temperature determined with the fit method (black circles) as well as via phase boundary estimates (blue circles) (e.g. from Fig. 3). Large changes in sample temperatures determined with the fit method are similar to those determined via phase boundary arguments.}
\end{figure}

The curves for which there is no shoulder-like feature can be fit to the functional form given in Eq. 2. The largest elastocaloric cooling response from the dataset displayed in Fig. 3(a) is fit in Fig. 4(a). Plotted alongside the fit is calculated temperature of the sample as a function of time from Eq. 1. The extracted value of the change in sample temperature, $\Delta T_s$, is approximately 2.36 K, nearly half of the sample temperature of 5 K, a very large cooling effect. Although the curves in Fig. 3(a) below 5 K cannot be adequately fit by Eq. 2, $\Delta T_s$ can still be estimated by realizing that elastocaloric cooling stops abruptly once the temperature of the sample reaches the phase transition temperature (i.e. the isentrope hits the solid black line in Fig. 3(b)) and also assuming that the thermometer is well thermalized with the sample. The estimation of $\Delta T_s$ for the dataset shown in Fig. 3(a) is plotted in Fig. 4(b) using both the fitting method for curves taken with $T\geq 5.1$ K and the intersection with phase boundary method for curves taken with $T < 5.1$ K. The change in sample temperature peaks near 5 K. More importantly, the relative order of magnitude of the quantities agree, ensuring that the methods are reasonable and that the sample is indeed cooling into the ordered state when the bath temperature is below 5 K.

\section{Discussion}

In a similar manner to what is done for room temperature elastocaloric cooling, figures of merit are important metrics by which materials can be compared. The most common figures of merit at room temperature are specific power (W/g) and power (W) \cite{Qian2024-RTCalorics}, which are based on temperature change and thermal \emph{resistance}. However, at cryogenic temperatures, minimizing the space taken by a cooling agent, and therefore maximizing its volumetric specific cooling power (W/cm$^3$), is imperative. The maximum volumetric specific cooling power, using the maximum change in temperature of the thermometer $\Delta T_t$ and the estimated thermal resistance (see Appendix F for more details) \cite{Hess2019-CaloricModeling} and converting via the density, is 0.34 W/cm$^3$ in our 0.19 mg sample of TmVO$_4$ at 5 K.

The specific cooling power, however, is an engineering-specific metric (since it relies on thermal resistance and not thermal resistivity); the geometry of the system will dictate it. In addition, energy efficiency is not as vital of a concern as is the ability to cool to the sub-Kelvin regime with little material. As an alternative, we introduce two material-specific metrics for elastocaloric materials used in cryogenic applications, the power gradient and energy density. The power gradient, in units of W/m, indicates how easily heat can be transferred to the cooling agent, and is calculated via the product of the inferred temperature change of the sample and the thermal conductivity, $\kappa_s \Delta T_s$. The energy density, in units of J/cm$^3$, demonstrates the energy reservoir present when cooling, and is taken to be the product of the inferred temperature change and the volumetric specific heat, $C_s\Delta T_s$. Using the reported thermal conductivity and heat capacity, the largest power gradient observed in the measurements presented here is 0.30 W/cm at 5 K, which has a corresponding energy density of $2.6 \cdot 10^{-3}$ J/cm$^3$. More information on the figures of merit can be found in Appendix F.

The theoretical entropy landscape, which was shown in Fig. 4(a) for qualitative purposes, can be compared against the inferred cooling shown in Fig. 4(a). Not including elastocaloric cooling from symmetry-preserving strain and assuming a rough symmetry decomposition of 90\% (see Appendix A), the expected cooling at 5 K given $\varepsilon^{(0)}_{x'x'} \approx -2.7 \cdot 10^{-3}$ and $\Delta\varepsilon_{x'x'} \approx 1.8\cdot 10^{-3}$ is approximately 3 K because the isentrope hits the phase boundary. Another more useful comparison is the amount of strain required to produce the same cooling, which is approximately $\Delta\varepsilon_{x'x'} \approx 1.3\cdot 10^{-3}$, indicating the realized elastocaloric cooling is of the correct order of magnitude as predicted by the Hamiltonian.

Although a large cooling effect was observed in this study, there is significant room for improvement in elastocaloric cooling at cryogenic temperatures. To further the use of quantum materials for elastocaloric cooling at cryogenic temperatures, both the technical and materials aspects of the problem must be addressed.

In principle, access to larger strains, both in terms of larger offset strains and larger changes in strains such that one has access to more points and paths in $\varepsilon-T$ space, will yield a larger elastocaloric effect. In addition, better thermalization with the device to be cooled (e.g. in this study, a thermometer), and less thermalization with the bath, will produce longer hold-times and a greater thermal transfer. Nevertheless, our measurements here serve as a proof-of-principle that even without these improvements, there is a material, or rather a class of materials, that can be explored and optimized to produce the most favorable cooling conditions.

Jahn-Teller materials, of which TmVO$_4$ is a specific realization, are a class of materials for which a large amount of entropy is preserved to low temperatures via degenerate crystal electric field levels. Ideally, the degeneracy can be lifted easily with strain because of direct, linear coupling between the electronic states and strain; the stronger this coupling, the larger the elastocaloric effect. Similar to adiabatic demagnetization, the best candidates for elastocaloric cooling to sub-Kelvin temperatures are those that remain paramagentic (or rather para\emph{nematic}) to 0 K and have rapidly growing (Curie-Weiss-like) susceptibility as $T\rightarrow 0$ K. This prevents isentropes from `running into' the ordered phase, which prevents further cooling. 

To search for candidate materials, 
AC elastocaloric effect, ultrasound, or any other method that are congruent to these measurements via Maxwell relations, are excellent; they are able to determine the strain susceptibility of a candidate materials. Spectroscopic probes including NMR \cite{Wang2021-NMRfluc,Vinograd2022-NMRZeeman,Nian2024-TmVO4NMRSpinEcho} and Raman scattering \cite{Ye2019-RamanYbRu2Ge2} that are also sensitive to rotational symmetry down to low temperatures are also insightful. Overall, however, measuring the AC elastocaloric effect is especially convenient, as the same tools can be used to perform the measurements presented in this study.

Aside from improvements, the AC elastocaloric effect has been shown to be an invaluable tool for better understanding quantum materials, and the strain load/unload technique can similarly be used. In a complementary way, the strain load/unload technique offers real-time information about the elastocaloric response of a material, which could reveal interesting relaxation behavior that may be difficult to observe otherwise.

\section{Conclusion}

In this study, a strain load/unload technique was used to measure the elastocaloric response of the Jahn-Teller insulator TmVO$_4$. The results are consistent with expectations based on the entropy landscape established from previous AC measurements \cite{Zic2024-TmVO4} and the known temperature dependence of the specific heat \cite{Massat2022-PNAS,Cooke1972-ortho,Li2024-TmYVO4} and thermal conductivity  \cite{Daudin1982-thermalConductivity,Vallipuram2024-thermalCond}. The data obtained provide direct realization of the predicted giant elastocaloric response that can otherwise only be inferred from AC techniques \cite{Zic2024-TmVO4}. This load/unload technique also clearly reveals the influence of the ordered phase and the changes in the time constant of the sample. Finally, the largest change in temperature of the sample was measured to be approximately 2.36 K at 5 K, a significant step in achieving viable cryogenic cooling using the elastocaloric effect. With rapid cycling, the absence of a magnetic field, and compactness, elastocaloric cooling using quantum materials, particularly Jahn-Teller materials, has tremendous potential as an alternative cooling method for cryogenic cooling applications.

\section*{Acknowledgments}

This work was supported by the National Science Foundation under grant no. DMR-2232515. MPZ was partially supported by a National Science Foundation Graduate Research Fellowship under grant number DGE-1656518. LY was partially supported by the Gordon and Betty Moore Foundation Emergent Phenomena in Quantum Systems Initiative through Grant GBMF9068.

\renewcommand{\figurename}{FIG.}
\renewcommand{\thefigure}{A\arabic{figure}}
\setcounter{figure}{0}

\section*{Appendix A: Experimental Setup and Detailed Methods}

As described in the main text, single crystals of TmVO$_4$ were grown using a molten flux technique. The single crystals were cut and polished, such that the [110] direction was the long-axis of the crystal, in a geometry optimal for strain measurements. The sample was suspended and epoxied between two pairs of titanium mounting plates with 2-ton and 5-minute epoxy, with the [110] axis aligned along the stress direction. A very thin layer of Angstrom Bond was added to the surface of the crystal to make the sample much less susceptible to cracking or breaking during the measurement. The sample plates were affixed to a CS100 Razorbill strain cell, which was used to control the strain that the material experiences via piezoelectric stacks. A thin piece of cigarette paper was used to prevent epoxy from flowing down the holes in the bottom sample plates.

A RuO$_2$ thermometer was attached to a gold heat pipe that was glued to the surface of the sample using DuPont 4929N silver paste prior to the application of Angstrom Bond, to allow for the largest possible thermal conductance between the thermometer and the TmVO$_4$ crystal. The heat-pipe thermally connects the thermometer to the sample to prevent any elastoresistance signal contamination in the elastocaloric response of the thermometer. The thermometer is electrically connected to a Wheatstone bridge in which each of the resistors in the four arms of the bridge comprise RuO$_2$ from the same batch to minimize variation. The bridge is affixed to the surface the strain cell, while thermally disconnected from the TmVO$_4$ crystal. This configuration allows for excellent cancellation of the large resistance of the thermometer attached to the sample, reducing the voltage input and enhancing signal-to-noise. In order to achieve even better cancellation due to slight mismatches in the values of the resistors, an adjustable resistor was added in parallel at room temperature to manually tune the background signal as close to zero as possible.

Fig. A1(a) illustrates a schematic of the experimental setup and Fig. A1(b) displays a picture of the mounted sample. Primed and unprimed axes refer to lab and crystal axes respectively. Specifically, unprimed $x$, $y$, and $z$ axes are aligned along the crystallographic [100], [010] and [001] axes. Since stress is applied along the [110] direction, a rotated (primed) coordinate system is used in the lab frame to describe the sample deformation. Hence, the $x'$ axis lies along the [110] crystallographic direction (see Fig. A1(c)).

\begin{figure}[!t]
\includegraphics[width=\linewidth]{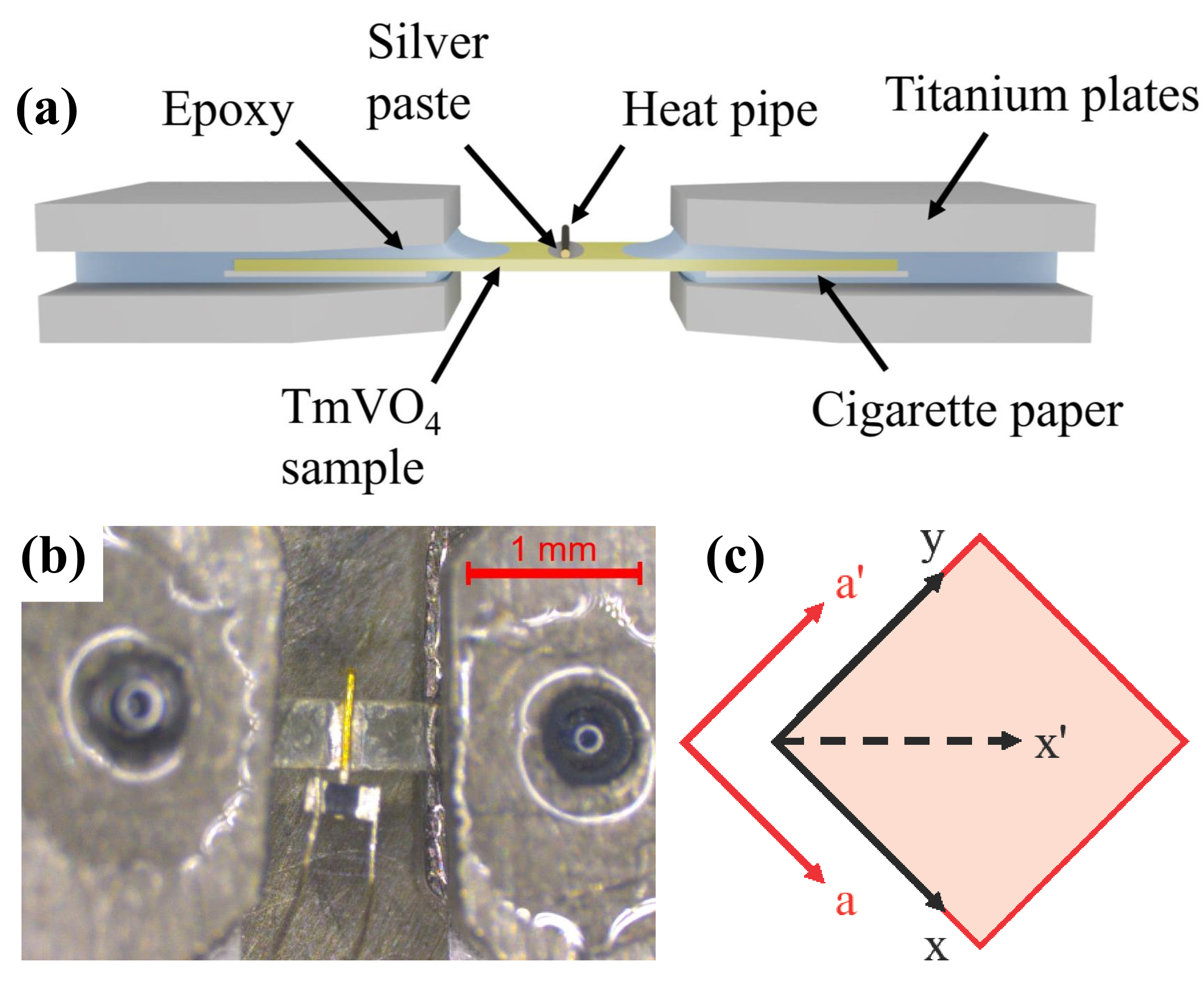}
\caption{(a) The sample (yellow, center) is suspended between two pairs of sample plates (gray) and held in place by epoxy (light gray). A heat pipe (center, brown) thermally connects the sample via silver epoxy to a thermometer (not shown). A thin layer of cigarette paper covers holes in the bottom sample plate to prevent epoxy from flowing down. The figure is a sketch and is not to scale. (b) The sample (center, transparent) is affixed to two pairs of sample plates (left and right) via epoxy, as discussed in the text. The temperature is measured via a RuO$_2$ thermometer (black film with two gold wire contacts). The thermometer is thermally connected to the sample via a gold heat pipe. All electrical and thermal connections on the thermometer are made with silver epoxy. (c) A different coordinate system relative to the principal crystal axes used for determining values of strain. The $x'$ direction is defined along the tetragonal [110] direction (dashed line), 45$\degree$ rotated from the crystallographic $a$ direction (red line), coincident with the $x$ direction (black line). }
\end{figure}

All measurements in this study were performed in a Quantum Design 14T PPMS using a custom-built strain-cell probe. The instrument setup for these experiments was identical to that used previously for AC elastocaloric measurements \cite{Ikeda2019-ECEtechnique}, except for a few key differences. First, as stated before, a RuO$_2$ thermometer was used to measure the temperature change of the sample instead of a thermocouple. Second, because the sample experienced pulses of strain rather than an oscillating strain, a function generator (SIGLENT SDG6022X Function Generator) was amplified by a high voltage amplifier (TEGAM Dual Channel High-Voltage Amplifier Model 2350) to generate the strain pulses. The strain pulse was sent to the outer stack of the Razorbill cell (varying between 40 V and 400 V), while the inner stack was controlled by a separate voltage source and remained constant during each pulse (usually $\pm 200$ V). The function generator was set to pulse mode with the fall edge set to 0.5 ms and the pulse width set to 200 ms. It was found empirically that a shorter fall edge produced a spurious voltage response in the TEGAM. Finally, the lock-in amplifier that measured the elastocaloric response of the material (SR860 Lock-In Amplifier) was configured much differently to previous AC measurements \cite{Ikeda2019-ECEtechnique}. It referenced an external frequency from the current source (100 $\mu$A, 1.453 kHz) to the Wheatstone bridge, had the sync filter enabled, and was configured with a time constant of 100 $\mu$s. To obtain the data, the Data Capture feature of the SR860 Lock-In Amplifier was used in one-shot mode with $n=4$, yielding a data acquisition rate of 20.3 kHz.

AC elastocaloric effect measurements were also performed, as detailed later in the appendices. These measurements were completed in the same manner as previous AC elastocaloric measurements in TmVO$_4$ \cite{Zic2024-TmVO4}, with a frequency of 223 Hz and peak-to-peak AC strain amplitude of $4.7 \cdot 10^{-5}$.

A capacitance sensor on the Razorbill CS100 Strain Cell was used to measure changes in separation across the cell as a proxy for changes in strain of the sample ($\varepsilon_{x' x'}$). The capacitance was read-out using a Keysight E4980AL LCR Meter. Zero strain was determined by examining the point at which the AC elastocaloric response arising from antisymmetric strain was zero; this was done at a sufficiently high temperature that effects arising from symmetry-preserving strain are negligible ($T > 8$ K) in the exact same manner as previous AC elastocaloric effect measurements on TmVO$_4$ (see supplementary material of ref. \cite{Zic2024-TmVO4}). The same finite element analysis strain transfer values, as had been determined alongside previous strain measurements on TmVO$_4$ (see supplementary of ref. \cite{Zic2024-TmVO4}), were used here. This temperature-dependent correction, which ranges from about 65\% to 90\% strain transfer, was included in the $\varepsilon_{x'x'}$ calculations throughout the manuscript. In principle, although not reported here, symmetry decomposition of $\varepsilon_{x'x'}$ to its symmetric ($(\varepsilon_{xx} + \varepsilon_{yy})/2$) and antisymmetric ($\varepsilon_{xy}$) components is also possible; this would be completed by transforming between unprimed and primed coordinates and using the known elastic moduli to determine the Poisson ratio (see supplementary of ref. \cite{Zic2024-TmVO4}). However, this decomposition is unnecessary here because we are interested in the total elastocaloric response arising from both symmetry strains.

\section*{Appendix B: Comparison of AC and strain load/unload technique results}

Both the strain load/unload technique (as shown in the main text) and the AC strain technique \cite{Ikeda2019-ECEtechnique} provide information on the time constants in the thermal model (Fig. 1(c) in the main text). To compare these two methods, AC elastocaloric measurements were also performed on the same sample used for strain load/unload measurements. Fig. A2(a) displays the AC elastocaloric response of the material as a function of frequency, also called the thermal transfer function, for different temperatures. In the thermal transfer function, the low-frequency tail and high frequency tail reveal $\tau_s$ and $\tau_t$, respectively. The magnitude of the thermal transfer function can be fit to the functional form \cite{Ikeda2019-ECEtechnique}:

\begin{equation}
    T_t(\omega) = \frac{T_{max}}{\sqrt{\Big(\frac{1}{\omega \tau_s} - \omega \tau_t \Big)^2 + \Big(1 + \frac{C_t}{C_s} + \frac{\tau_t}{\tau_s}\Big)^2}}.
\end{equation}
Fig. A2(a) also includes the fits to the above functional form. The shape of the thermal transfer function changes as the temperature is lowered towards the phase transition; a characteristic `plateau' feature emerges because thermal transfer is maximized.

To compare the two methods, Fig. A2(b) shows the obtained values of $\tau_s$ from both the AC elastocaloric effect and the strain load/unload technique used in the main text as a function of temperature. The agreement is reasonable, establishing similar temperature dependence and corroborating the two measurements.

\begin{figure}[!t]
\includegraphics[width=\linewidth]{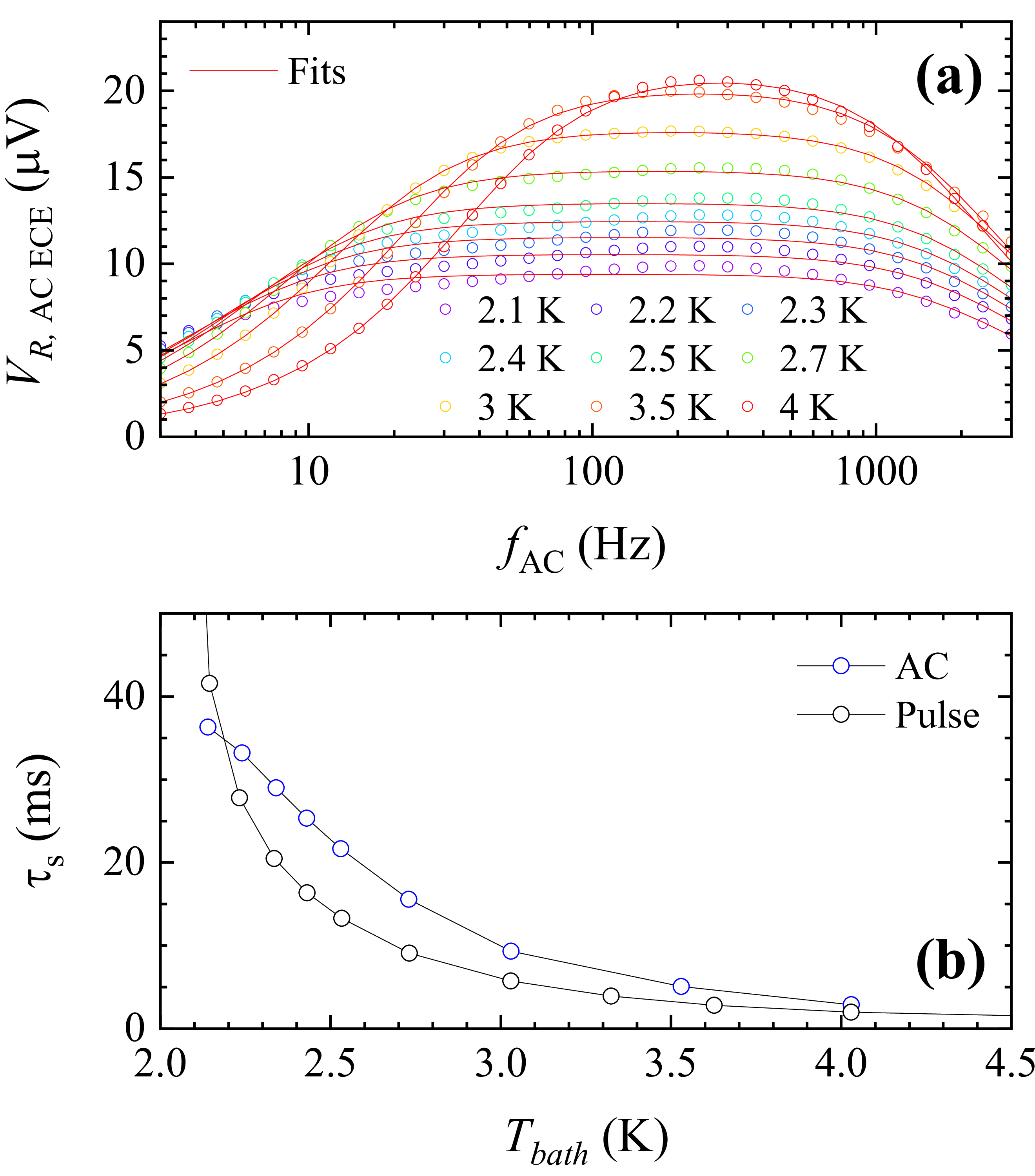}
\caption{(a) Frequency dependence of the magnitude of the AC elastocaloric response. Different colored points correspond to different temperatures. Solid red lines correspond to fits to Eq. 3. (b) From the fits performed in panel (a) to the frequency dependence of the AC elastocaloric effect and in the main text to the load/unload elastocaloric effect data, the time constant of the sample is plotted as a function of temperature.}
\end{figure}

\section*{Appendix C: Heating from piezoelectric stacks and background subtraction}

\begin{figure}[!t]
\includegraphics[width=\linewidth]{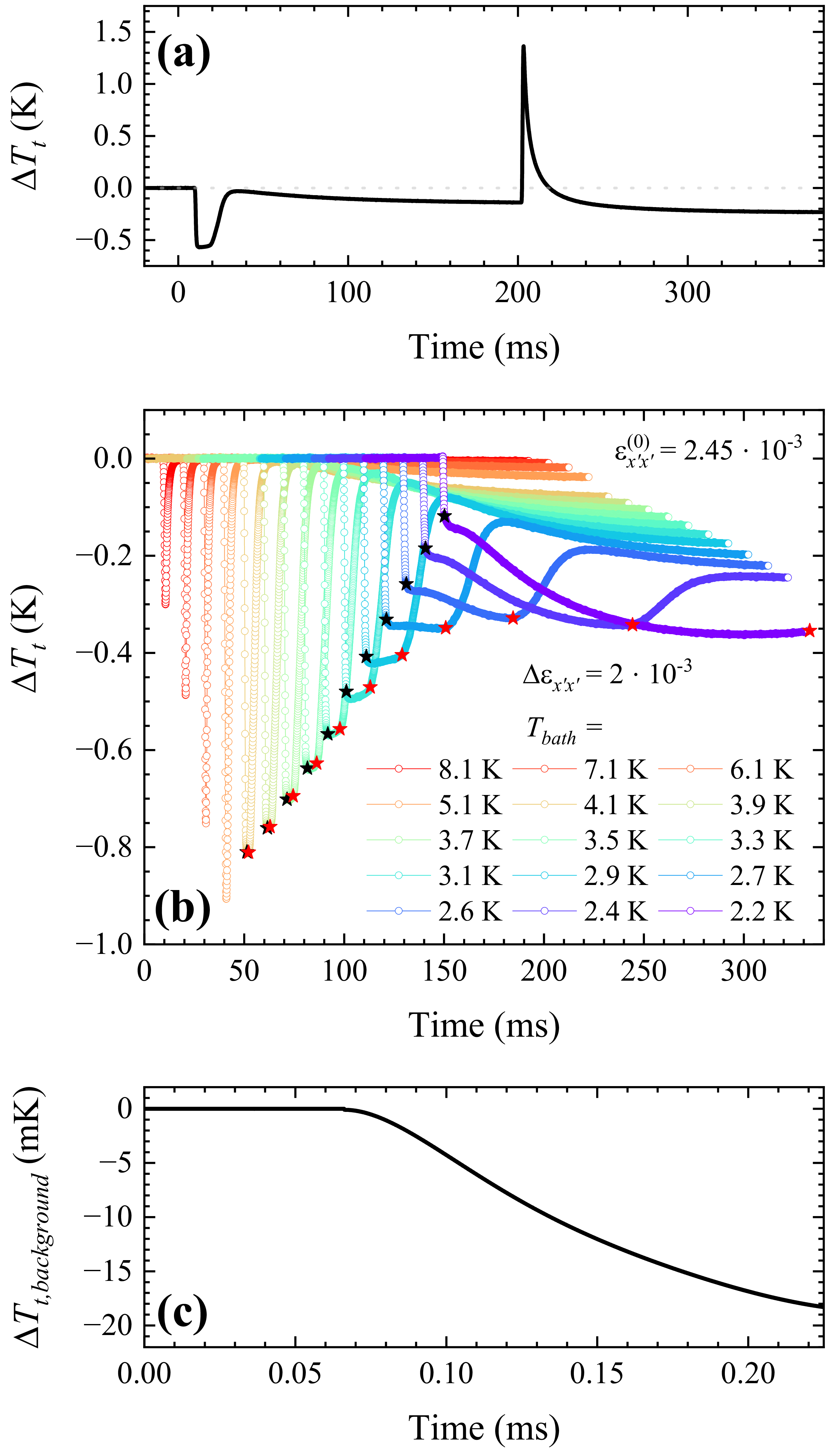}
\caption{(a) A full load/unload elastocaloric signature as measured by the thermometer as a function of time. After the strain load phase, a signature of cooling over a long period of time (between 30 and 200 ms) appears after the sample has thermally relaxed. Then, after the strain unload phase, the sample appears to cool even further (e.g., compare the values at 150 ms and 350 ms). (b) The same representative elastocaloric response as measured by the thermometer ($\Delta T_t$) as a function of time (in milliseconds) under a large strain pulse ($\Delta \varepsilon_{x'x'} = 1.8 \cdot 10^{-3}$) for different bath temperatures ($T_{bath}$) with a initial strain of $\varepsilon_{x'x'}^{(0)} = -2.45 \cdot 10^{-3}$ as shown in Fig. 3(a) of the main text but with the long-time background feature present. Once again, black stars label the entrance to, and red stars label the exit from, the plateau region.  (c) The long-time curve from the 7.1 K data that was scaled and subtracted from the other curves in panel (b) to produce Fig. 3(a).}
\end{figure}

As established earlier in the appendices, the strain the sample experienced was controlled by applying a potential difference across piezoelectric stacks ($-200$ to $200$ V). These piezoelectric stacks can be thought of as effective capacitors, and, under a sudden change in voltage, heat with a finite power. Specifically, the power dependence of an ideal capacitor is:

\begin{equation}
    P(t) = V(t) \frac{dV(t)}{dt} C.
\end{equation}
With significant changes in the potential (up to $\pm 400$ V in this work), the piezoelectric stacks may undergo non-negligible increases in temperature. 

Fig. A3(a) shows the full time dependence of an elastocaloric response using the largest changes in strain used for this study. There was a spurious, low-amplitude response that persisted for a long time. This response also possessed the same sign for both the load and unload portions of the strain pulse. Here, piezoelectric stack heating of the Wheatstone bridge, which is used to measure the change in temperature of the sample, manifested as a cooling signature in the elastocaloric response.

Fig. A3(b) displays the same figure as Fig. 3(a) in the main text but without background subtraction. Although the heating response varied as a function of temperature, all but the amplitude appeared to remain the same. So, to more clearly present the shoulder-like feature, the long-time response of the 7 K curve, as shown in Fig. A3(c), was used to remove the background from all of the curves by scaling and then subtracting to finally obtain Fig. 3(a) in the main text. Note that the background subtraction does not change the location of the shoulder-like feature and solely removes the broad, extrinsic response.

\section*{Appendix D: Thermal time constants and their influence on the elastocaloric signature}

\begin{figure}[!t]
\includegraphics[width=\linewidth]{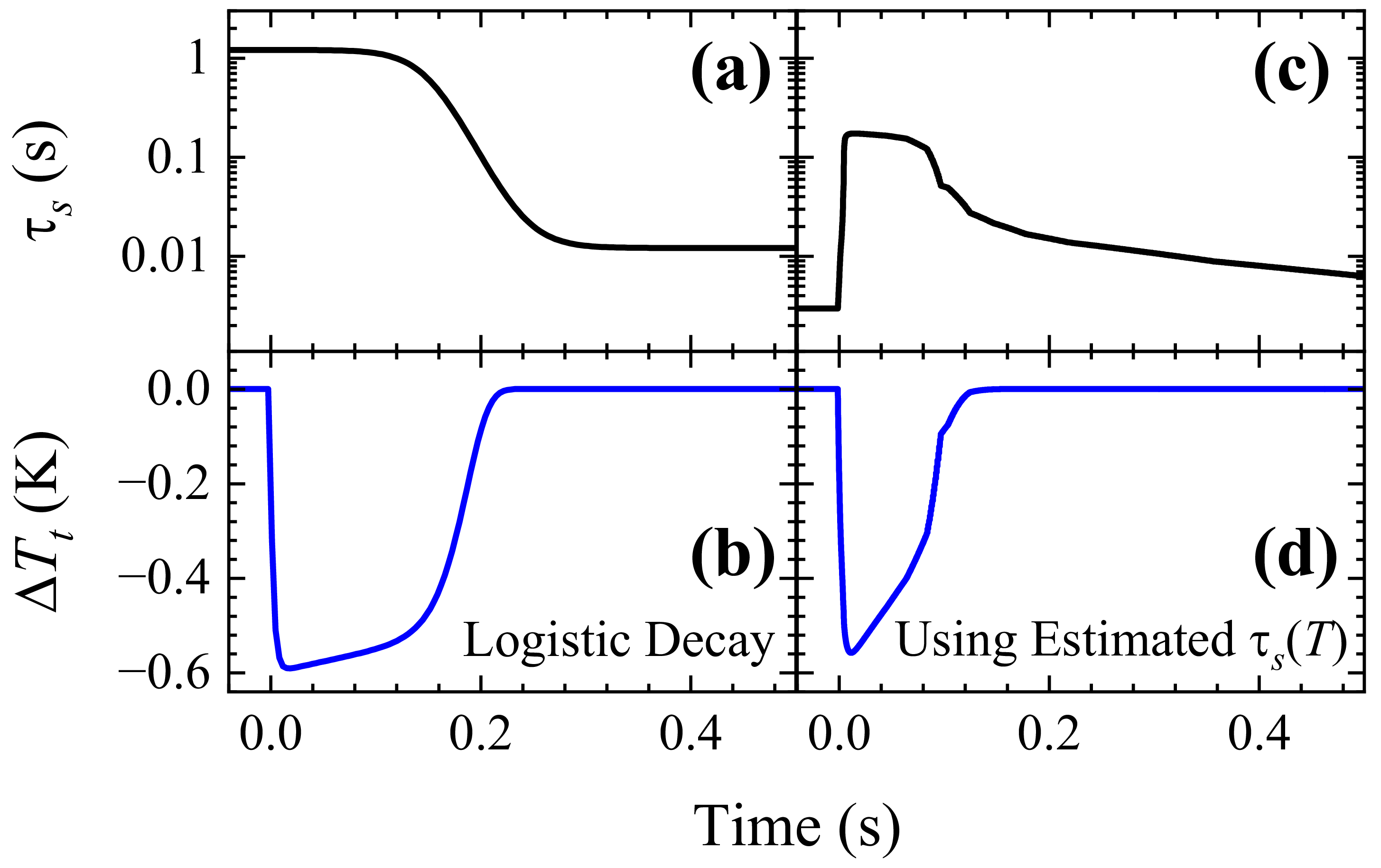}
\caption{Variation of $\tau_s(t)$ and $\Delta T_t$ for a logistic decay model (a,b) and using the estimated $\tau_s(T)$ shown in Fig. 3(c) of the main text (c,d), respectively. }
\end{figure}

In an effort to reproduce the shoulder-like feature in Fig. 3(a) in the main text, different methods were used. A first \emph{empirical} attempt was made by assuming that the sample time constant, as a function of time, behaved with a functional form described by logistic decay as shown in Fig. A4(a):
\begin{equation}
    f(t) = \frac{C}{1 + e^{-k(t - t_0)}},
\end{equation}
where, $C$, $k$, and $t_0$ are independent parameters. This method was intended to model the qualitative behavior of the time constant of the sample. Using this and Eq. 2 from the main text, the calculated, convolved elastocaloric response is displayed in Fig. A4(b) (using $\tau_t = 3.1\cdot 10^{-3}$ s, $\tau_s = 1.2 \cdot 10^{-2}$ s, and $\Delta T_s = 0.6$ K). This method of calculation captured much of the qualitative behavior of the shoulder-like feature.

Another attempt was made using the time constant as a function of temperature shown in Fig. 3(c) in the main text. This method used the temperature of the material as feedback for the selected time constant. In this model, after an initial change in temperature, the sample relaxed and the time constant was adjusted based on the temperature at that time. This then produced the time constant as a function of temperature, shown in Fig. A4(c), and the resulting elastocaloric response, in Fig. A4(d). This method did not produce as flat a feature as is in the data. This could be because of inaccuracies in the estimation of the thermal time constant of the sample, which is the quotient of two experimentally determined quantities.

\begin{figure}[!t]
\includegraphics[width=\linewidth]{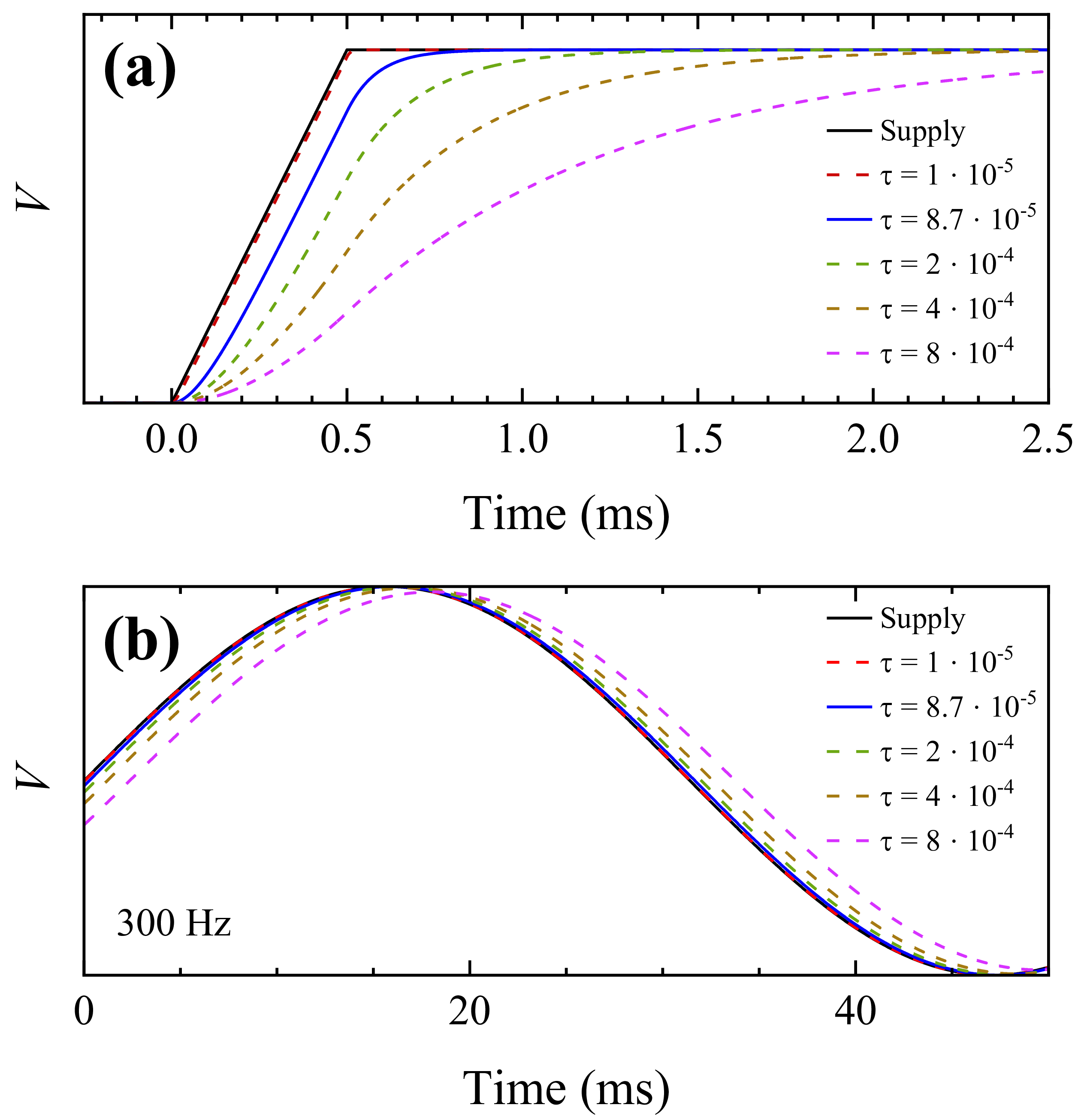}
\caption{In both panels, the solid blue line represents the potential across the piezoelectric stacks under the experimental conditions present in the setup used to conduct the measurements for this study and the solid black line represents a voltage source. Other colored, dashed lines represent the potential across the capacitor different RC time constants for illustration. Panel (a) shows the potential across the piezoelectric stacks for the load/unload measurement for a rise time of 500 $\mu$s, while panel (b) shows the potential across the piezoelectric stacks for the AC elastocaloric effect measurements with a frequency of 300 Hz.}
\end{figure}

\section*{Appendix E: Circuitry time constant}

Considering a simple circuit with a resistor, capacitor, and voltage source, the potential across the capacitor depends on the RC time constant, $\tau_{RC}$, present. Given that the potential supplied to the piezoelectric stacks is a pulse with amplitude $A$ and has a rise time of $w$, the charge $Q$ on the capacitor satisfies:
\begin{equation}
Q'_{\text{PZT}}(\omega)R + \frac{Q_{\text{PZT}}(\omega)}{C} = 
\begin{cases} 
0 & t < 0 \\
At & 0 \leq t < w \\
Aw & t \geq w
\end{cases}.
\end{equation}
Solving this for $Q$ and computing the potential across the piezoelectric stacks, the solution is:
\begin{equation}
V_{\text{PZT}} = 
\begin{cases} 
0 & t < 0 \\
ACR \left( e^{-\frac{t}{RC}} - 1 \right) + At & 0 \leq t < w \\
ACR \left( 1 - e^{-\frac{w}{RC}} \right)e^{-\frac{t}{RC}} + Aw & t \geq w
\end{cases}.
\end{equation}
The supplied voltage and potential across the piezoelectric stacks are plotted in Fig. A5(a) for different time constants. The time constant corresponding to the experimental setup used in this study is $\tau = 8.7 \cdot 10^{-5}$ s. Even for the largest changes in voltage supplied, the RC time constant is short enough such that there is no significant change in strain within a meaningful time window, albeit a small time difference between the supplied and actual potential. At the time scales and temperatures present in these measurements, creep and hysteresis are not expected \cite{Islam2018-PZTthesis}. No other effects are expected aside from those arising from electromechanical resonance, which would lead to a spurious voltage response and was not evident in our measurements.

The same can also be done for an AC supply, which is pertinent for the AC elastocaloric measurements performed. The charge on the capacitor is given by:
\begin{equation}
V'_{\text{PZT}}(\omega)R + \frac{V_{\text{PZT}}(\omega)}{C} = \sin \omega t,
\end{equation}
and the potential across the capacitor is:
\begin{equation}
V_{\text{PZT}} = \frac{\sin \omega t}{C^2 R^2 \omega^2 + 1} - \frac{CR \omega \cos \omega t}{C^2 R^2 \omega^2 + 1}.
\end{equation}
For frequencies near the one used for the AC elastocaloric measurements presented here (223 Hz), there is a negligible change in the waveform across the piezoelectric stacks for $\tau = 8.7 \cdot 10^{-5}$ s, as shown in Fig. A5(b).

\section*{Appendix F: Figures of merit calculations}

The volumetric specific cooling power (0.34 W/cm$^3$) was calculated by using the maximum value of the cooling power stated in the literature \cite{Hess2019-CaloricModeling} along with the value of $\Delta T_t$ at 5 K (0.90 K) and the thermal resistance of the sample (obtained with the dimensions of the sample and the thermal conductivity given in the literature \cite{Daudin1982-thermalConductivity}):

\begin{equation}
    \dot Q = \frac{\Delta T}{2 \pi R_{therm}},
\end{equation}
and subsequently dividing by the mass of the sample (0.19 mg sample), the thermal resistance ($R_{therm}$) and converting with the density of TmVO$_4$ (6.02 g/cm$^3$ \cite{Chakoumakos1994-TmVO4density}). 

The power gradient and energy density were calculated by taking the product of $\Delta T_s$ at 5 K (2.36 K) and the thermal conductivity \cite{Daudin1982-thermalConductivity} and the specific heat (see supplementary material of ref. \cite{Zic2024-TmVO4}), respectively.
\bibliography{bib}

\end{document}